\documentclass[preprintnumbers,amsmath,amssymb,11pt,a4paper]{revtex4}
\usepackage{amsmath,amsfonts,amssymb,graphics,graphicx,epsfig,subfigure,color,bbm}

\begin{document}

\title{\textbf{\large{Oscillations in the total photodetachment cross sections of a linear triatomic negative ion}}}

\author{A Afaq}\email{draafaq@gmail.com} \affiliation{Center of Excellence in Solid State Physics, University of the Punjab, Quid-e-Azam Campus, Lahore-54590, Pakistan}
\author{M. Hanif, Iftikhar Ahmad}\email{ahma5532@gmail.com}\affiliation{Department of Physics, Hazara University, Mansehra,
Pakistan}

\begin{abstract}
The total photodetachment cross section of a linear triatomic
negative ion using plane polarized laser light parallel to the
axis of the molecular ion is derived. The cross section shows
strong oscillations in contrast to the recently reported case,
where the laser polarization was perpendicular to the axis of the
triatomic molecular ion (Appl. Phys. Lett. {\bf{94}}, 041125,
2009). Closed orbit theory is used to explain these oscillations
and the results are compared with a diatomic molecule.
\vspace{.5cm}\noindent
\end{abstract}
\maketitle

\section{Introduction}
Interest in the photodetachment process and
photodetachment spectroscopy has gown in the recent years
\cite{afaq09,du10,Afaq2009,Afaq2008,Afaq2006,Biv08,bergues07,Bergues,Wells,walter,anderson,pern}.
The photodetachment cross section for hydrogen negative ion has
been investigated experimentally \cite{A3,A4} and theoretically
\cite{Du2004,Du2006,A5,A6,A64,A7,A72,A73,A74}. The most intrusting
aspect of the H$^-$ cross section in the presence of a static
electric field of a few hundred kV/cm above threshold, it shows
oscillations while smooth in the absence of the field. Afaq and Du
extended one-center model and developed two-center model for the
photodetachment of electron from a diatomic molecule
\cite{Afaq2006,Afaq2008,Afaq2009}. They gave the idea of special
detached-electron orbit starting from one atom and ending at the
other atom in a diatomic negative molecule. This special
detached-electron orbit is responsible for the oscillations in the
total photodetachment cross sections and is the updated version of
the closed orbit theory \cite{A64, du1988,d1988}.

Recently Afaq $et~ al$ \cite{afaq09} showed a smooth behavior of
the total photodetachment cross section of a linear triatomic
negative ion placed perpendicular to the direction of polarization
of the laser beam. The total cross section approaches one-center
and two-center models for a highly energetic laser beam while it
approaches to one-center model for a low energy laser. In the
present article we have investigated the photodetachment cross
section for a linear triatomic negative ion with the laser beam
polarized parallel to the axis of the triatomic ion. A procedure
similar to the one reported by Afaq $et~ al$ \cite{afaq09} is used
in the present calculation.

Three centers of the linear triatomic negative ion is supposed as
a coherent source of electrons similar to H$^-$ and the
detached-electron wave function is obtained as a superposition of
the three coherent waves originating from each center of the
system. The detached-electron flux, due to the superposition of
these three waves, on a screen is calculated. For the total
photodetachment cross section, of a linear triatomic negative ion,
an analytical formula is derived by the integration of the
electron flux in all directions. Our present article will be
helpful in the understanding of the structural information of a
linear triatomic negative ions like $BeCl_2^-,HCN^-,CS_2^-,CO_2^-$
etc. \cite{Bhardwaj1997}. Atomic units are used throughout the
article or mentioned otherwise.

\section{Detached electron wave function}
The schematic diagram of the linear triatomic negative ion as a
three-center model is shown in Fig.1. Numbers 1, 2 and 3 on the
$z$ axis represent the three centers of the system for the
negative molecular ion, such that number 3 is set as the origin of
the coordinates system. The $\pm$ symbols indicate the sign of the
two lobes of the p-orbital wave function. The screen is placed at
a distance $L$ from the three-center system and perpendicular to
the z axis. Assume $d$ is the distance between the two adjacent
centers and is of the order of a few atomic units, while $L$ is
much larger than $d$ and is equal to several thousands atomic
units \cite{A73,A74}. These two variables can be changed in our
model.

Let there is only one active electron like the H$^-$ model
\cite{A6}. The normalized wave function for the active electron in
the three-center system is the linear superposition of the H$^-$
like bound state at the three centers,
$\phi_T=(\phi_1+\phi_2+\phi_3)/\sqrt3$. Where $\phi_1$, $\phi_2$
and $\phi_3$ are the normalized wave functions for H$^-$ but
centered at 1, 2 and 3 in Fig. 1. Photodetachment process is a two
step process \cite{B3}. In the first step, a negative ion absorbs
one photon energy $E_{ph}$ and generates an outgoing electron
wave, while in the second step, this outgoing wave propagates to a
large distance.

A z-polarized laser light similar to Ref. \cite{afaq09} is used.
The detached-electron wave function can be obtained by the linear
superposition of the three coherent waves generated from the each
center. These coherent waves can be achieved from the results of
the H$^{-}$ photodetachment in the absence of an electric field
\cite{A7}. Assume $\Psi^+_{1}$, $\Psi^+_{2}$ and $\Psi^+_{3}$ are
the waves emitted from centers 1, 2 and 3. The detached electron
wave function $\Psi^{+}_M$ is given by:

\begin{equation}
  \Psi^{+}_M=\frac{1}{\sqrt{3}}(\Psi^+_{1}+\Psi^+_{2}+\Psi^+_3)
\end{equation}
Spherical polar coordinates $(r_1,\theta_1,\phi_1)$,
$(r_2,\theta_2,\phi_2)$ and $(r_3,\theta_3,\phi_3)$ for the
detached-electron relative to each center are used. The emitted
waves, $\Psi^+_{1}$, $\Psi^+_{2}$ and $\Psi^+_{3}$, can be
expressed using Ref. \cite{A7} by:
$$\Psi^+_{1}(r_{1},\theta_{1},\phi_{1})=C\cos\theta_{1}\frac{\exp(ikr_{1})}{kr_{1}}$$
$$\Psi^+_{2}(r_{2},\theta_{2},\phi_{2})=C\cos\theta_{2}\frac{\exp(ikr_{2})}{kr_{2}}$$
$$\Psi^+_{3}(r_{3},\theta_{3},\phi_{3})=C\cos\theta_{3}\frac{\exp(ikr_{3})}{kr_{3}},$$

where $C=\frac{4kBi}{(k_{b}^{2}+k^{2})^{2}}$, $k_{b}$ is related
to the binding energy $E_{b}$ of H$^-$ by
$E_{b}=\frac{k^2_{b}}{2}$, and $B$ is a normalization constant,
having value of $0.31552$.

The detached-electron wave function $\Psi^{+}_M(r,\theta,\phi)$
from the three-center system can be obtained after the
substitution of each center wave function in Eq. (1). The problem
can be further simplified by using the idea that $L$ is much
larger than $d$. Let $(r,\theta,\phi)$ be the spherical
coordinates of the detached electron.  For the phase terms, $
r_{1}\approx r-d\cos\theta$, $ r_{2}\approx r+d\cos\theta$ and
$r_3\approx r$ are used, while in all other places $r_{1}\approx
r_{2} \approx r_{3}\approx r$ and $\theta_{1}\approx \theta_{2}
\approx \theta_{3}\approx \theta$. With these approximations
$\Psi^{+}_M(r,\theta,\phi)$ reduces to:

\begin{equation}
\Psi^{+}_M(r,\theta,\phi)=\frac{C\cos\theta} {\sqrt{3}}
\Big[1+2\cos(kd\cos\theta)\Big]\frac{\exp(ikr)}{r}
\end{equation}

\section{Derivation of the total cross section for the triatomic negative ion}
The detached-electron flux in the radial direction for the
triatomic negative ion is calculated as in \cite{afaq09}:

\begin{equation}
{j_{r}}(r,\theta,\phi)=\frac{kC^{2}\cos^{2}\theta}{3r^{2}}\Big[1+4\cos(kd\cos\theta)+
4\cos^2(kd\cos\theta)\Big]
\end{equation}

A large imaginary spherical surface $\Gamma$ enclosing the
triatomic negative ion is used to investigate the behavior of the
total cross section. From the electron flux crossing the surface a
generalized differential cross section $\frac{d\sigma(q)}{ds}$ can
be obtained \cite{A7}:

\begin{equation}
\frac{d\sigma(q)}{ds}=\frac{2\pi
E_{ph}}{c}\vec{{j_{r}}}\cdot\hat{n}
\end{equation}

where q is the coordinate on the surface $\Gamma$, $\hat{n}$ is
the exterior norm vector at q, c is the speed of light in a.u. and
$ds=r^{2}\sin\theta d\theta d\phi$ is the differential area on the
spherical surface. The total cross section of the triatomic
negative ion can be derived by the integration of the differential
cross section over the entire surface,
$\sigma(q)=\int_{\Gamma}\frac{d\sigma(q)}{ds}ds.$  Substituting
the value of $\vec{{j_{r}}}$ from Eq. (3) in Eq. (4) and
integrating:
$$\sigma(E)=\frac{2\pi k
C^{2}E_{ph}}{3c}\int_{0}^{\pi}\int_{0}^{2\pi}\cos^2\theta\sin\theta[1+4\cos(kd\cos\theta)+
4\cos^2(kd\cos\theta)]  d\theta d\phi$$

After solving the integration, the following simple result is
obtained:
\begin{equation}
\sigma(E)=\sigma_{0}(E)A_{||}(kd)
\end{equation}
where
\begin{equation}
\sigma_{0}(E)=\frac{8\pi^{2}k |C|^{2}E_{ph}}{3 c}~~~~~ with~~~~~
C=\frac{4kBi}{(k_{b}^{2}+k^{2})^{2}}
\end{equation}

\begin{equation}
A_{||}(kd)=1+4\Big[\frac{\sin(kd)}{(kd)}+2\frac{\cos(kd)}{(kd)^2}-2\frac{\sin(kd)}{(kd)^3}\Big]
+2\Big[\frac{\sin(2kd)}{(2kd)}+2\frac{\cos(2kd)}{(2kd)^2}-2\frac{\sin(2kd)}{(2kd)^3}\Big]
\end{equation}

$\sigma_{0}(E)$ is the smooth total photodeatchment cross section
of $H^-$ \cite{Du1988} and $A_{||}(kd)$ is the modulation function
due to the triatomic negative ion. The total photodetachment
cross-section expressed in Eq. (5) is plotted in Fig. 2 as a
function of the incident laser photon enenrgy (eV) for several $d$
values. The figure shows strong oscillation for the laser
polarization parallel to the axis of the linear triatomic
molecule, in contrast to the recently reported \cite{afaq09} laser
polarization perpendicular to the triatomic molecule. Closed orbit
theory \cite{A64,du1988,d1988} is used to explain the origin of
these oscillations. The detached-electron wave produced at center
$1$ while propagates out of the source reaches to center $3$ and
center $2$ on its way out. The overlap of the detached-electron
wave from center $1$ with the source at center $3$ and the source
at the center $2$ produces oscillations in the total cross section
of the triatomic negative ion. As a result we encounter two
special orbits of the detached-electron; one is generated from
center $1$ to the nearby center $3$ and the other one is from
center $1$ to center $2$. These two special orbits of the
detached-electron are responsible for the oscillations. Similar
oscillations are also reported in the diatomic negative ion
(two-center system) \cite{Afaq2009}. A comparison of our results
with the two-center model is shown in Fig. 3. Fourier
transformation in Fig. 3 is used to investigate special orbits of
the detached-electron.

\section{Conclusion}
A plane polarized laser, polarized parallel to
the axis of the linear tri-atomic negative ion, is used to
investigate the behavior of the detached electron from a triatomic
negative ion. A simple analytical formula for the total cross
section is obtained which shows strong oscillations in contrast to
the case reported in \cite{afaq09}. Closed orbit theory is used to
explain these oscillations as the interference between
detached-electron waves produced from one center and sources at
the other two centers. Fourier transformation in Fig. 3 shows that
these two types of orbits are responsible for the oscillations in
the total photodetachment cross section and hence there are two
oscillation frequencies in the three-center system in contrast to
only one oscillation frequency in the two-center system. Our
theoretical study suggests that the photodetachment of negative
ion as a three-center model is interesting, and the interference
can be used to investigate structural information of linear
triatomic negative ions. However this model can optimally be
modified for non-linear triatomic ions.

\clearpage
\begin{center}
{\bf Figure captions}
\end{center}

\begin{description}
\item[Fig.1] Schematic diagram of the three-center model as a
triatomic negative ion
\end{description}

\begin{description}
\item[Fig.2] The total photodetachment cross section for different
values of $d$ using Eqs. (5)-(7). From (a) to (d) show that with
the increase in $d$ the oscillation amplitude decreases and
oscillation frequency increases
\end{description}

\begin{description}
\item[Fig.3] Fourier transformation of the total photodetachment
cross section (a) calculation, (b) comparison: the dashed line is
the result with two-center system and solid line is the result
with three-center system. The calculations are carried out using
$d=250$ a.u.
\end{description}


\begin{thebibliography}\small{
\bibitem{afaq09} Afaq A, Ahmad  I, Ahmad M A, Rashid A, Tahir B A, and Hussain M T 2009 Appl. Phys. Lett. {\bf{94}} 041125
\bibitem{Afaq2009} Afaq A and Du M L 2009 J. Phys. B: At. Mol. Opt. Phys. {\bf{42}} 105101
\bibitem{du10} Yang B C and Du M L 2010 J. Phys. B: At. Mol. Opt. Phys. {\bf{43}} 035002
\bibitem{Afaq2008} Afaq A and Du M L 2008 Commun. Theor. Phys. {\bf{50}} 1401
\bibitem{Afaq2006} Afaq A and Du M L 2006 Commun. Theor. Phys. {\bf{46}} 119
\bibitem{Biv08} Bivona S, Bonanno G, Burlon R, Gurrera D and Leone C 2008 Phys. Rev. A {\bf{77}} 051404(R)
\bibitem{bergues07} Bergues B, Ansari Z, Hanstorp D and Kiyan I Y 2007 Phys. Rev. A {\bf{75}} 063415
\bibitem{Bergues} Bergues B and Kiyan I Y 2008 Phys. Rev. Lett. {\bf{100}} 143004
\bibitem{Wells} Wells J E and Yukich J N 2009 Phys. Rev. A {\bf{80}} 055403
\bibitem{walter} Walter C W, Gibson N D, Field R L, Snedden A P, Shapiro J Z, Janczak C M and Hanstorp D 2009 Phys. Rev. A {\bf{80}} 014501
\bibitem{anderson} Andersson P, Lindahl A O, Hanstorp D and Pegg D J 2009 Phys. Rev. A {\bf{79}} 022502
\bibitem{pern} Pernpointner M, Rapps T and Cederbaum L S 2009 J. Chem. Phys. {\bf{131}} 044322
\bibitem{A3} Bryant H C, Mohagheghi A H, Stewart J E, Donahue J B, Quick C R, Reeder R A, Yuan V, Hummer C R, Smith W W, Cohenl S, Reinhardt W P and Overman L 1987 Phys. Rev. Lett. {\bf{58}} 2412
\bibitem{A4} Stewart J E, Bryant H C, Harris P G, Mohagheghi A H, Donahue J B, Quick C R, Reeder R A, Yuan V, Hummer C R, Smith W W and Cohenl S 1988 Phys. Rev. A {\bf{38}} 5628
\bibitem{Du2004} Du M L 2004 Phys. Rev. A {\bf{70}} 055402
\bibitem{Du2006} Du M L 2006 Eur. Phys. J. D {\bf{38}} 533
\bibitem{A5} Rau A R P and Wong H 1988 Phys. Rev. A {\bf{37}} 632
\bibitem{A6} Du M L and Delos J B 1988 Phys. Rev. A {\bf{38}} 5609
\bibitem{A64} Du M L and Delos J B 1987 Phys. Rev. Lett. {\bf{58}} 1731
\bibitem{A7} Du M L 1989 Phys Rev A {\bf{40}} 4983
\bibitem{A72} Fabrikant I I 1990 J Phys. B:  At. Mol. Opt. Phys. {\bf{23}} 1139
\bibitem{A73} Blondel C, Delsart C and Dulieu F 1996 Phys. Rev. Lett. {\bf{77}} 3755
\bibitem{A74} Blondel C, Chaibi W, Delsart C, Drag C, Goldfarb F and Kr$\ddot{o}$ger S 2005 Eur. Phys. J. D {\bf{33}} 335
\bibitem{du1988} Du M L and Delos J B 1988 Phys. Rev. A. {\bf{38}} 1896
\bibitem {d1988} Du M L and Delos J B 1988 Phys. Rev. A. {\bf{38}} 1913
\bibitem{Bhardwaj1997} Bhardwaj V R, Mathur D and Rajgara F A 1998 Phys. Rev. Lett. {\bf{80}} 3220
\bibitem{B3} Bracher C, Delos J B, Kanellopoulos V, Kleber M and Kramer T 2005 Phys. Lett. A {\bf{347}} 62
\bibitem{Du1988} Du M L and Delos J B 1988 Phys. Rev. A {\bf{38}} 5609



 }

\end{thebibliography}
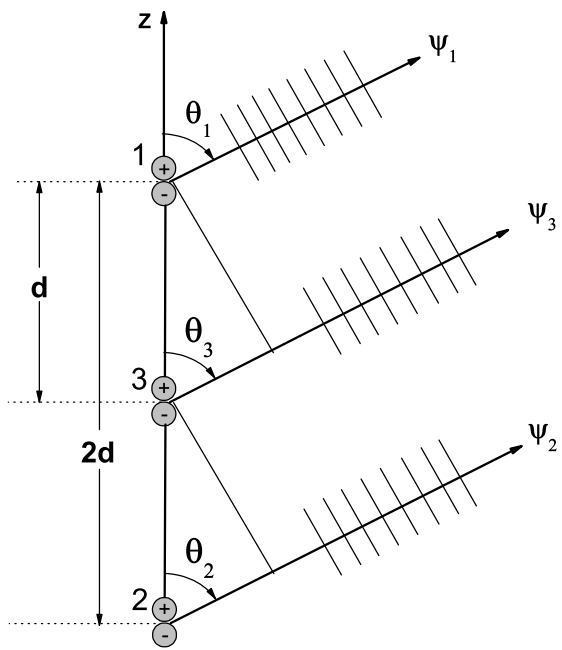
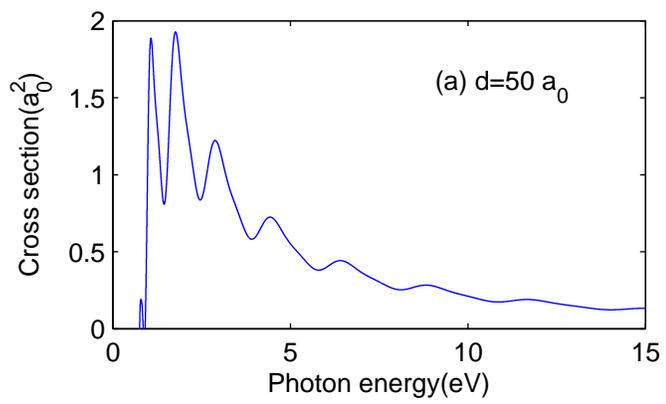
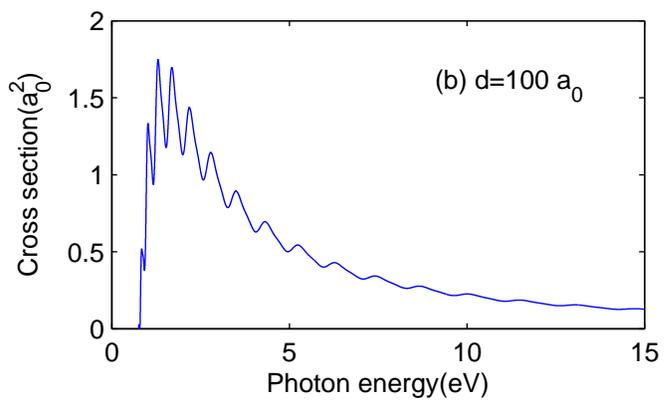
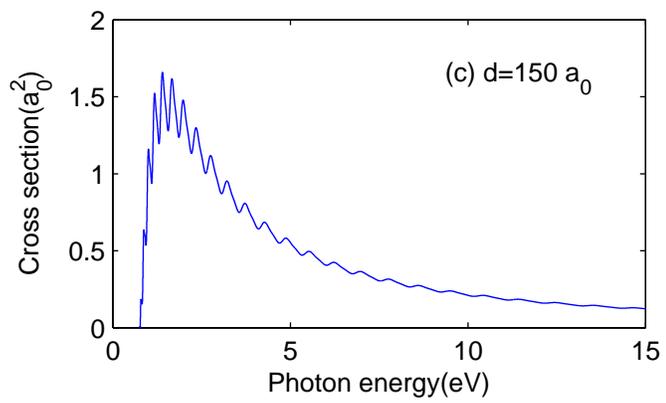
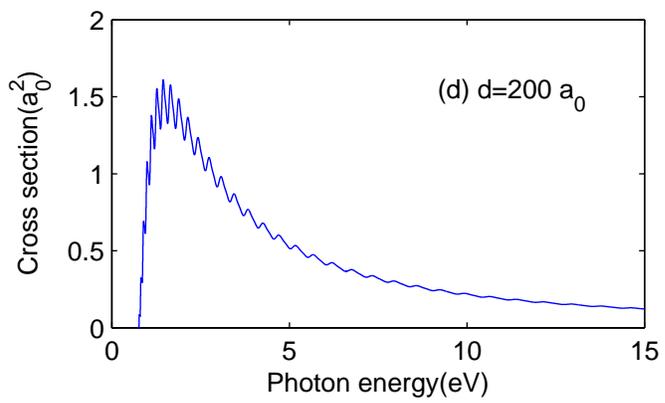
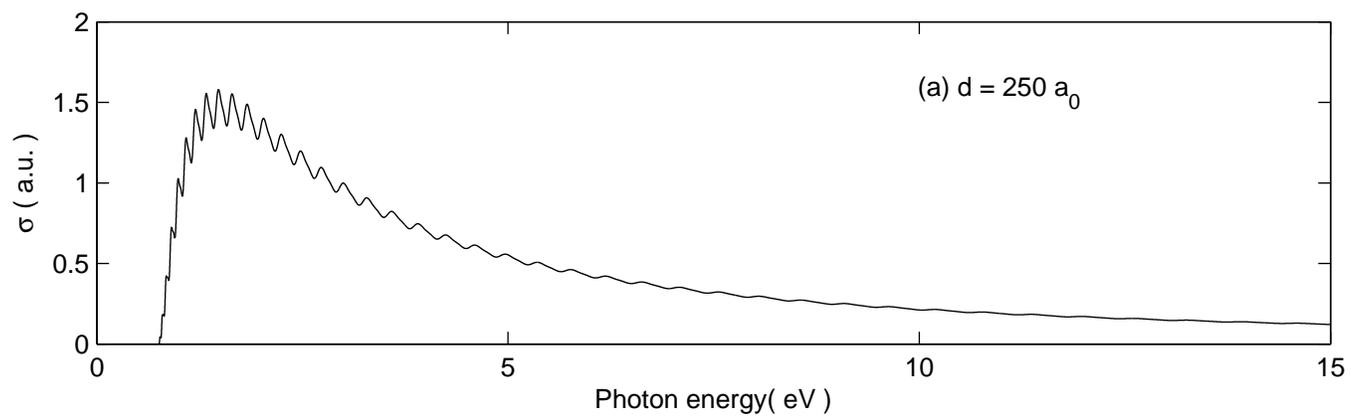
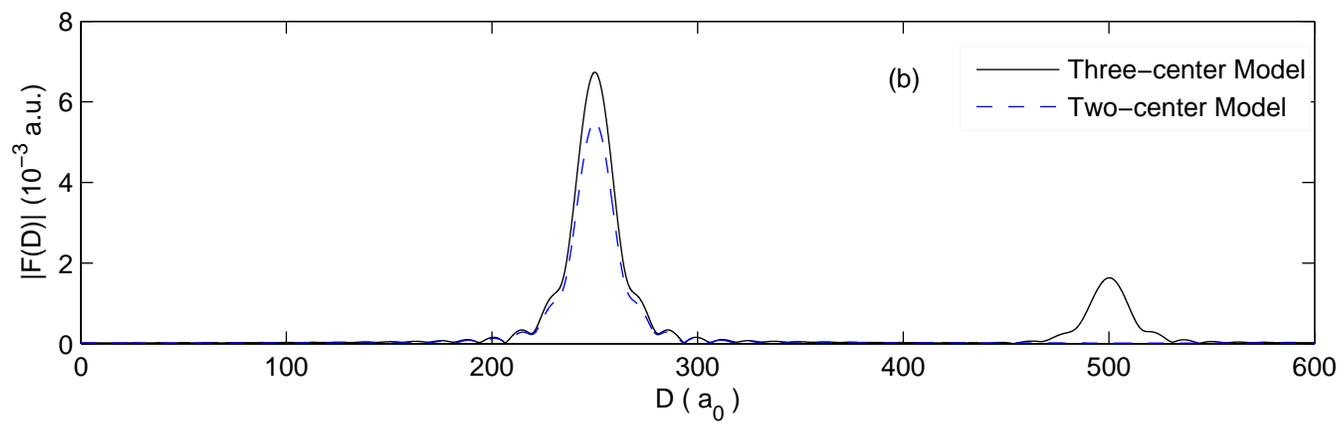
\end{document}